# Nonlinear nanophotonic devices in the Ultraviolet to Visible wavelength range

Jinghan He[1], Hong Chen[2], Jin Hu[3], Jingan Zhou[2], Yingmu Zhang[4], Andre Kovach[4], Constantine Sideris[3], Mark C. Harrison[5], Yuji Zhao[2], Andrea M. Armani[1, 3, 4, *]

[1] Dept of Chemistry, University of Southern California, Los Angeles, CA 90089 USA

[2] School of Electrical, Computer, and Energy Engineering, Arizona State University, Tempe, AZ, 85287 USA

[3] Ming Hsieh Dept of Electrical and Computer Engineering, University of Southern California, Los Angeles, CA 90089 USA

[4] Mork Family Dept. of Chemical Engineering and Materials Science, University of Southern California, Los Angeles, CA 90089 USA

[5] Fowler School of Engineering, Chapman University, Orange, CA 92866 USA

**Abstract**

Although the first lasers invented operated in the visible, the first on-chip devices were optimized for near-infrared (IR) performance driven by demand in telecommunications. However, as the applications of integrated photonics has broadened, the wavelength demand has as well, and we are now returning to the visible (Vis) and pushing into the ultraviolet (UV). This shift has required innovations in device design and in materials as well as leveraging nonlinear behavior to reach these wavelengths. This review discusses the key nonlinear phenomena that can be used as well as presents several emerging material systems and devices that have reached the UV-Vis wavelength range.

## 1 Introduction

The past several decades has witnessed the convergence of novel nonlinear materials with nanofabrication methods, enabling a plethora of new nonlinear optical (NLO) devices(1–4). Originally, the focus was on developing devices operating in the telecommunications wavelength band to improve communications. One example of an initial success is on-chip modulators and add-drop filters for switching and isolating of optical wavelengths. While initial devices were fabricated from crystalline materials (5–7), the highest performing devices were made from organic polymers (8–16). As nanofabrication processes improved, higher performance integrated devices were developed, such as high quality factor optical resonant cavities, and higher order nonlinear behaviors became accessible. This technology enabled on-chip frequency combs (2, 17, 18), stokes and anti-stokes lasers (19–21), and super continuum sources (22).

While these devices can be used in many fields, one clear application of these devices is in quantum optics. While many quantum phenomena can be investigated using near-infrared (IR) lasers, atomic clocks based on Rb and Ce transition lines require visible lasers as excitation sources. Initial work developing proof of concept systems relied on large optical lasers. More recently, the focus shifted to "clocks on a chip" (23–26). Because the transition lines are in the visible, the development of an ultra-narrow linewidth and stable visible laser source at a complementary wavelength was a key stepping stone. Similarly, over the past few years, a plethora of novel quantum emitters have been discovered (27). However, the majority are excited in the visible. In order to realize integrated devices based on these new materials, it is necessary to have an integrated source with sufficient power.

Visible sources also play a key role in biotechnology, namely, the miniaturization of diagnostics and imaging systems. Tissue and biosamples absorb strongly in the near-IR wavelength range. This absorbance will degrade the performance of many diagnostic techniques, and it can result in scattering and signal degradation in imaging (28–30). In addition, many imaging methods require fluorescent probes. The majority of light emitting bio-labels are excited in the ultraviolet (UV) to visible wavelength range (~300 - ~700nm) (31–33). Therefore, to support this rapidly emerging field, there is a growing effort to develop complementary integrated sources.

This review will introduce the key theoretical mechanisms that underpin nonlinear interactions in integrated photonic devices. These act as design rules for both the devices discussed here as well as devices in general. Then, a discussion of several new crystalline and organic materials and devices being actively used to achieve ultraviolet (UV)-visible (Vis) emission with be reviewed. Lastly, a discussion of possible new research directions will be presented.

## 2 Basics of Nonlinear Optics

### 2.1 Background

Maxwell's equations form the basis for describing electric and magnetic fields at a macroscopic level. Combining them, one can obtain the wave equation which is the foundation for electromagnetic radiation, also known as light. Two of Maxwell's equations for electric displacement (D) and magnetic field (H) are given below:

$$\nabla \cdot \mathbf{D} = \rho_f \tag{1}$$

$$\nabla \times \mathbf{H} = \mathbf{j}_f + \frac{\partial \mathbf{D}}{\partial t} \tag{2}$$

where $\rho_f$ is the free charge density, and $j_f$ is the free current density. Furthermore, **D** and **H** can be obtained through the constitutive relationships. We will focus on **D**, because most materials (and the materials discussed herein) are nonmagnetic, so **H** is directly related to **B** via $\mu_0$. The electric displacement, **D**, is related to the electric field, **E**, via the permittivity ($\varepsilon_0$) and the polarization density (**P**):

$$D = \epsilon_0 E + P \tag{3}$$

Furthermore, the polarization can be represented by a sum of its linear and nonlinear parts: $P = P^{(L)} + P^{(NL)}$ (4). Assuming a plane-wave propagating in the z direction with amplitude *A*, angular frequency *ω*, and propagation vector *k*, we use the wave equation to relate the field amplitude to nonlinear polarization. Applying the slowly-varying approximation (the field varies slowly with propagation distance), which applies in most nonlinear materials, and assuming negligible loss we arrive at:

$$\frac{dA(\omega)}{dz} = j \frac{\omega}{2n\epsilon_0 c} P^{(NL)}(\omega) e^{-jkz} \tag{5}$$

In other words, the amplitude of the field as a function of frequency will vary depending on the nonlinear polarization density, which is medium dependent. The particular form of the nonlinear component of polarization will depend on the nonlinear process generated in the material, but in general the polarization density can be expanded in a power series. The first term represents the linear part, and all subsequent terms represent the nonlinear part:

$$P = \epsilon_0 \chi^{(1)} E + \epsilon_0 \chi^{(2)} E^2 + \epsilon_0 \chi^{(3)} E^3 + \ldots \tag{6}$$

The χ terms represent different orders of nonlinear susceptibility, and each is a tensor with terms to mix the x, y, and z components of the electric field. Additionally, equation 6 is only valid in the frequency domain or for ultrafast nonlinearities in non-dispersive materials. In other instances in the time domain, overlap integrals with the response time are required, but for our simple analysis, equation 6 will suffice. We will explore specific components of nonlinear polarization in the following sections and discus how they can be leveraged to generate frequencies in the UV–Vis range.

## 2.2  $\chi^{(2)}$ effects: Second harmonic generation and three wave mixing

In this section, we focus on the second-order terms, which correspond to the $\chi^{(2)}$ expansion term, and are thus often referred to as $\chi^{(2)}$ effects. The $\chi^{(2)}$ coefficient is often represented mathematically with the d coefficient ($d = \frac{1}{2}\chi^{(2)}$). Since both of these quantities tensors, they include terms to mix the x, y, and z components. The $\chi^{(2)}$ effects collectively include several 3-wave mixing effects: second-harmonic generation (SHG or frequency doubling), difference-frequency generation, and sum-frequency generation to name a few. For simplicity, we treat these effects together. For $\chi^{(2)}$ effects, we first consider three frequencies of light traveling through a crystal (Figure 1a) such that $\omega_3 = \omega_1 + \omega_2$. Now we examine the coupled wave equations that result from nonlinear polarization density being a tensor and consider only the frequencies that satisfy the enforced constraint $\omega_3 = \omega_1 + \omega_2$. The resulting coupled wave equations are:

$$\frac{d}{dz}E_1(z) = -j\omega_1 \sqrt{\frac{\mu_0}{\epsilon_1}} dE_3\, E_2^* e^{-j(k_3-k_2-k_1)z} \tag{7}$$

$$\frac{d}{dz}E_2(z) = -j\omega_2 \sqrt{\frac{\mu_0}{\epsilon_2}} dE_3\, E_1^* e^{-j(k_3-k_2-k_1)z} \tag{8}$$

$$\frac{d}{dz}E_3(z) = -j\omega_3 \sqrt{\frac{\mu_0}{\epsilon_3}} dE_2\, E_1\, e^{-j(k_3-k_2-k_1)z} \tag{9}$$

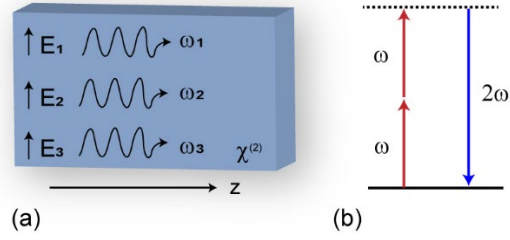

(a) (b)

Figure 1: (a) Three waves with frequencies $\omega_1$, $\omega_2$, and $\omega_3$ mix in a nonlinear optical crystal exhibiting 2nd-order ($\chi^{(2)}$) effects, propagating in the z-direction. (b) For second-harmonic generation (SHG), two photons at the pump frequency are destroyed while interacting with the material to produce a single photon at double the pump frequency (half the wavelength).

For propagation through thin materials, we assume that the pump waves ($\omega_1$ and $\omega_2$) are not depleted, therefore $\frac{d}{dz}E_1(z) = \frac{d}{dz}E_2(z) = 0$. We integrate equation (9) over z from 0 to L (with L being the length propagated through the nonlinear material) and assume $E_3(0)=0$ ($\omega_3$ is the frequency being generated). Substituting the **E** fields for intensity ($I = \frac{n}{2\sqrt{\frac{\mu_0}{\epsilon_0}}} EE^*$) we can write the optical intensity as:

$$I_3(L) = 2\left(\frac{\mu_0}{\epsilon_0}\right)^{\frac{3}{2}} \frac{(\omega_3 dL)^2}{n_1 n_2 n_3} I_1 I_2 \left[\frac{\sin(\frac{\Delta k L}{2})}{\frac{\Delta k L}{2}}\right]^2$$

(10)

Where

$$\Delta k = k_3 - k_2 - k_1 = [\omega_3 n_3 - \omega_2 n_2 - \omega_1 n_1]/c \tag{11}$$

n is the frequency-dependent index of refraction, and c is the speed of light in vacuum. It is important to notice that the intensity will vary with a regular beat length due to the presence of the sine function. This regular sinusoidal variation of intensity (or power) over distance is sometimes referred to as power cycling, and it limits nonlinear conversion efficiencies. We will discuss how to address that with phase matching in a subsequent section, but for now, let us consider a special case for the chosen frequencies. For frequency doubling (SHG), we will let $\omega_1 = \omega_2 = \omega$ and $\omega_3 = 2\omega$ and reconsider equation 10. In the process of SHG, two photons at frequency $\omega$ are destroyed to create a single photon at frequency $2\omega$ (Figure 1b).

$$I_{2\omega}(L) = 2(\frac{\mu_0}{\epsilon_0})^{\frac{3}{2}} \frac{(\omega dL)^2}{n_\omega^2 n_\omega} I_\omega^2 sinc^2\left[\frac{\Delta kL}{2}\right] \tag{12}$$

In order to let the intensity of $\omega_3$ grow beyond what is limited by the sinc function (that is, beyond the maximum allowed by the periodic variation over distance), we must get $\Delta k=0$, and the way to achieve that is to match the phase of the mixed waves. This is conventionally done by taking advantage of the anisotropic nature of most nonlinear optical materials, but in nanophotonics in particular, it can also be done via modal phase matching or dispersion engineering. The anisotropic nature means that the permittivity, refractive index, and therefore propagation vector *k* are tensors and will be a function of the polarization of the **E** field. The mechanics of phase matching are not discussed here, but we examine it briefly in section 2.4. With proper alignment of the optical fields with the nonlinear medium and phase matching, the sinc function goes to 1 and the equation reduces to

$$I_{2\omega}(L) = 2(\frac{\mu_0}{\epsilon_0})^{\frac{3}{2}} \frac{(\omega dL)^2}{n_\omega^2 n_{2\omega}} I_\omega^2$$

(13)

Although it is possible to get high conversion efficiencies with SHG in crystals, it can be difficult, so this model is sufficient for understanding the nonlinear behavior (34–37). At high enough intensities, the pump energy will deplete and the frequency-doubled output will start to diminish and will eventually saturate. It is also important to note that, due to the phase matching condition, the polarization of the second-harmonic frequency will typically be orthogonal to the polarization of the pump frequency. For Type 0 phase matching, described in more detail in section 2.4, the phase matching is not achieved via anisotropy but instead by modal phase matching, quasi-phase matching, or dispersion engineering, and the second harmonic frequency can have the same polarization as the pump frequency (38). Finally, we note that the modal overlap of the pump and SHG frequencies is important for SHG (or any 2nd-order process) to occur. Mode overlap is trivial for beams within bulk nonlinear materials, but is an important consideration for nanophotonic devices (39, 40).

2nd-order nonlinear effects, and especially SHG, are frequently used for converting IR light sources to UV–Vis range. For example, common green laser pointers are often made by frequency-doubling a 1064 nm wavelength source to 532 nm light using a nonlinear crystal. Although nonlinear conversion efficiencies are low, this scheme is often more efficient than creating a laser that emits at 532 nm directly due to the high efficiency of IR lasers, particularly those that emit at 1064 nm.

## 2.3   $\chi^{(3)}$ effects: Third harmonic generation and four wave mixing

In the previous section, we considered the second-order term in the polarization power series (equation 6). In this section, we consider the 3rd-order term, $\chi^{(3)}$. Instead of considering three-wave mixing, we now consider four-wave mixing interactions in a nonlinear medium. In general, one can analyze four separate wavelengths mixing (Figure 2a). For simplicity, one will often analyze the degenerate case where $\omega_1 = \omega_2 = \omega_3 = \omega_4 = \omega$. We can generate four coupled-wave equations which can be useful for phase-conjugate mirrors, but not for upconverting IR light. However, the treatment extends to third-harmonic generation (THG), which is analogous to second-harmonic generation.

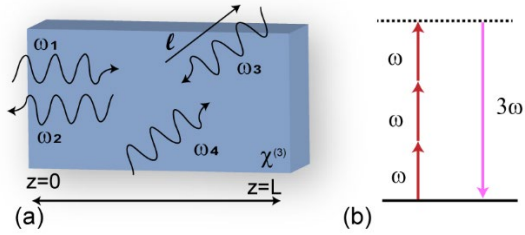

Figure 2: (a) Four waves with frequencies $\omega_1$, $\omega_2$, $\omega_3$, and $\omega_3$ mix in a nonlinear optical crystal exhibiting 3rd-order ($\chi^{(3)}$) effects, with two waves propagating in the z-direction and one propagating along vector $\ell$. (b) For third-harmonic generation (THG), three photons at the pump frequency are destroyed while interacting with the material to produce a single photon at triple the pump frequency (1/3 the wavelength). In THG, the incident (pump) photons and third-harmonic photons can propagate in the same direction.

In THG, three incident photons of frequency $\omega$ are destroyed to create a single photon at frequency $3\omega$ (Figure 2b). This process is useful for upconverting IR light into the visible, or more often UV range. We can follow a similar analysis for SHG and calculate the THG intensity ($I_{3\omega}$):

$$I_{3\omega}(L) = \frac{\left(3\omega\chi^{(3)}\right)^2}{16\epsilon_0^2 c^4 n_{3\omega} n_\omega^3} L^2 I_\omega^3 sinc^2\left[\frac{\Delta k L}{2}\right] \quad (14)$$

where $I_\omega$ is the fundamental beam intensity and L is the distance propagated through the material. We see again the presence of a sinc function dependent on $\Delta k$, meaning we have similar phase-matching concerns as with 2nd-order nonlinearities to avoid problematic power cycling. In bulk materials, phase matching is often more difficult for THG as the material is not always anisotropic. The availability of other phase matching techniques, such as modal phase matching, make nanophotonics an ideal platform for studying THG and other $\chi^{(3)}$ effects. Just as with 2nd-order effects, the third-harmonic and fundamental frequencies must have modal overlap with the nonlinear material and each other in order for THG to take place.

Practically speaking, THG via $\chi^{(3)}$ effects is inefficient. In many cases of THG, a second harmonic is generated via $\chi^{(2)}$ which then mixes with the unconverted frequency in another $\chi^{(2)}$ interaction (sum-frequency generation), producing the third harmonic, which tends to be more efficient than the $\chi^{(3)}$ process by itself. However, both have been used to demonstrate upconversion, as shown in references (37, 41–46). Additionally, there are some important applications of $\chi^{(3)}$ processes, such as the generation of frequency combs (17, 47–55). A frequency comb is a source with a spectrum that contains a series of equally-spaced frequency lines.

Another $\chi^{(3)}$ effect is the Kerr effect, in which high-intensity light induces a refractive index change in the material (given by $n = n_0 + n_2 I$), which results in self-phase modulation. The Kerr effect contributes to other $\chi^{(3)}$ effects, such as dispersive wave generation, which is the result of soliton dynamics. In dispersive wave generation, a soliton propagating with a frequency distance $\delta\omega$ from the zero-dispersion wavelength coherently couples to an optical wave with a frequency distance -$2\delta\omega$ from the zero-dispersion wavelength, which will be in-phase with the soliton. The zero-dispersion wavelength is the wavelength at which the material dispersion and modal dispersion offset one another. Dispersive wave generation can be analyzed as a cascaded four-wave mixing process (56).

Two-photon absorption (TPA) is a nonlinear process that occurs in a material where a single photon does not have enough energy to span the gap between a ground state and an excited state. At high enough intensities and when the energy gap to the excited state is equivalent to the energy of two photons, both photons can be simultaneously absorbed by the material, resulting in TPA. It is important to note that TPA is distinct from SHG, as SHG involves the conversion of two photons to a higher-frequency photon and TPA involves the absorption of two photons into a real excited state of the material. TPA can be modeled as a $\chi^{(3)}$ process using the following differential equation

$$\frac{dI}{dz} = -\alpha I - \beta I^2 \tag{15}$$

where $\alpha$ is the linear loss and $\beta$ is the two-photon absorption coefficient. At the high intensities needed to drive other nonlinear processes, TPA can be a significant source of loss, especially in the UV–Vis range.

## 2.4 Phase matching in nonlinear materials (and quasi-phase-matching)

Phase matching for $\chi^{(2)}$ processes is conventionally achieved by using an anisotropic crystal where you can have different refractive indices based on polarization. Because $\chi^{(2)}$ nonlinearities are only available in non-centrosymmetric crystals, they can take advantage of this anisotropy. Fewer $\chi^{(3)}$ materials are anisotropic so phase matching is often achieved by other means. Depending on the nonlinear crystal, there are two ways to achieve phase matching in uniaxial crystals, as shown in Table 1. In a practical sense, this means that for SHG and THG using this phase matching, the upconverted wavelength will be at a different polarization than the pump wavelength.

Table 1: Strategies to achieve phase matching in uniaxial crystals

|  | Positive Uniaxial ($n_e > n_o$) | Negative Uniaxial ($n_e < n_o$) |
|---|---|---|
| Type I | $n_3^o \omega_3 = n_1^e \omega_1 + n_2^e \omega_2$ | $n_3^e \omega_3 = n_1^o \omega_1 + n_2^o \omega_2$ |
| Type II | $n_3^o \omega_3 = n_1^o \omega_1 + n_2^e \omega_2$ | $n_3^e \omega_3 = n_1^e \omega_1 + n_2^o \omega_2$ |

When phase matching may not be possible (as in a nonlinear waveguide), quasi-phase matching is often employed. In this method, the optical axis of the nonlinear material is made to alternate at regular intervals (Figure 3). This eliminates the problem of power cycling by inverting the phase with respect to the axis at a specified period, $\Lambda$. Creating the periodic optic axis in the structure is usually achieved via poling the material: using a very high electric field to force the axis to align with the electric field lines (57).

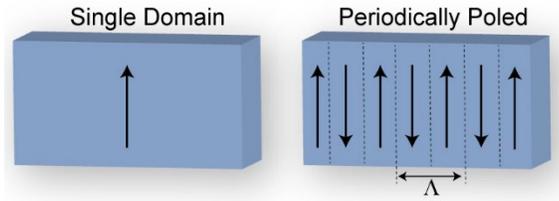

Figure 3: A crystal with a single domain (left) will have a single optical axis. With periodic poling (right) the crystal will have alternating optic axis with a period of Λ. Periodically-poled materials are frequently used for phase-matching in nonlinear interactions.

While phase matching is important for nonlinear interactions, particularly SHG and THG, it is also important to note that for thin films (short interaction lengths) it is less crucial. In integrated optical systems where there is some other condition imposed on the orientation of the crystal or isotropic nonlinear materials are used, phase matching conditions must often be met using other strategies. These other phase matching conditions (including quasi-phase matching) are often referred to as Type 0 phase matching. One example is using modal dispersion to phase-match, which may depend on material choices and/or the specific geometry of the device (58, 59). In modal dispersion phase matching, the device is designed such that the effective refractive index of the modes supporting the pump and generated (second-harmonic or third-harmonic) frequencies match. Additionally, the dispersion of the light signal may be engineered (via the use of metamaterials or photonic crystals, for example) in order to ensure phase matching conditions are met. These other phase matching techniques are particularly important, because in an integrated device, it is often not possible to rotate the nonlinear crystal axes with respect to the optical field once the device has been fabricated, and isotropic nonlinear materials are attractive for applications in the UV–Vis range.

## 2.5  Other effects: Raman scattering, Brillouin scattering, Supercontinuum

Raman scattering is a process where a photon traveling through a material loses energy by exciting vibrational states of the material or gains energy from those same vibrational states. In a typical Raman process, energy from the photon is lost to the material vibrational modes, creating a longer wavelength referred to as a Stokes wavelength or Stokes shift. Once vibrational modes are resonating, subsequent pump photons may pick up energy from the vibrational mode, creating an anti-Stokes wavelength or anti-Stokes shift (Figure 4). The relationship of scattered power ($P_{scatter}$) to the incident intensity is characterized by a scattering cross-section ($\sigma_R$), which is typically measured experimentally:

$$P_{scatter} = \sigma_R I_0 \qquad (16)$$

where $I_0$ is the incident intensity. Typically, Stokes scattering alone is not useful for upconverting IR to UV or visible light (it downconverts or redshifts the wavelength), but anti-Stokes scattering (which occurs in conjunction with Stokes scattering) can be used for upconversion. This is typically efficient in systems where there is a very large intensity, such as high-Q resonators where photons circulate with a very long lifetime, or high confinement, as in plasmonic nanoparticles or structures (60–64).

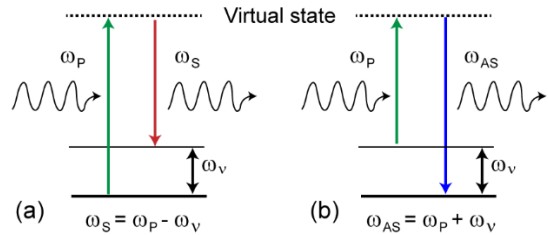

Figure 4: Stimulated Raman scattering can generate Stokes and anti-Stokes frequencies. (a) For a Stokes shift, energy from the incident photon is lost to vibrational modes of the material and a lower frequency (longer wavelength) photon is scattered. (b) For an anti-Stokes shift, the incident photon gains energy from the vibrational modes of the material and the scattered photon has a higher frequency (shorter wavelength).

Stimulated Raman scattering is a process where the pump and Stokes wavelengths are introduced into the material to generate an enhanced Raman response where energy is transferred from the pump wavelength to the Stokes wavelength. The interactions from stimulated Raman scattering are typically analyzed as $\chi^{(3)}$ processes with the Stokes intensity given by:

$$I_s(L) = I_s(0) e^{g_R I_P L} \qquad (17)$$

where $g_R$ is the Raman gain intensity factor, $I_S$ is the Stokes beam intensity, $I_P$ is the pump beam intensity, and $L$ is the propagation length for the undepleted pump approximation. The Raman gain intensity factor is often measured experimentally, although it can be defined in terms of refractive indices for the pump and Stokes beam, the Stokes frequency and the imaginary part of $\chi_R^{(3)}$, the Raman susceptibility. Because the pump and Stokes beams are coupled via the vibrational modes in the material, Raman scattering is always phase-matched in the propagation direction of the pump beam.

As opposed to exciting vibrational states of the material (Raman scattering), photons may also excite acoustic waves within a material, which will generate periodic refractive index variations (the refractive index depends on density). If a photon is subsequently scattered off this acoustic wave, the result is Brillouin scattering. Stimulated Brillouin scattering is a process where a spontaneously scattered photon with a frequency difference from the pump equal to the frequency of the acoustic wave couples with the pump beam to drive the acoustic wave. This most often occurs when the photon scatters in the backward direction and can be a limiting factor in fiber optic communications. Stimulated Brillouin scattering has recently seen much interest in the integrated photonics community, and it is discussed in more detail in reference (65).

One important application of nonlinear phenomena is the generation of supercontinua. A supercontinuum is a light signal with a broad and smooth spectrum. They are generated by multiple nonlinear processes interacting, such as four-wave mixing, Raman scattering, sum-frequency generation, dispersive wave generation, and self-phase modulation (a process where light intensity induces a refractive index change in the material, leading to a chirped pulse). The specific nonlinear effects used for generating a supercontinuum will depend on the materials being used and occasionally on the geometry of the device used for generation. Supercontinuum generation has been used across the visible and UV range in nanophotonic devices (66, 67).

## 3 Nonlinear Optical Materials

### 3.1 Materials overview

In the past a few years, photonics based on narrow bandgap semiconductors such as Si and GaAs III-V have achieved great success in nanophotonic devices as well as integrated photonics circuits (PICs). However, the narrow bandgap energy of these materials, e.g., 1.1 eV of Si, has restricted the light transmission to wavelengths longer than 1130 nm, which has hindered their applications in UV–Vis spectral region. Alternatively, wide bandgap semiconductor/dielectric materials, such as silica, SiN, SiC, diamond, lithium niobate, and III-N (e.g., GaN and AlN), have received considerable attention for the nanophotonic applications in the UV–Vis range.

Silicon dioxide (or silica) possesses the largest bandgap energy among these materials, thus exhibiting extreme broadband transparency in the visible through the near-IR. Additionally, over the past several decades, numerous wet chemistry and deposition routes have been developed for making both undoped and doped silica layers (68–72). When combined with the ease of processing and the compatibility of silica with common micro-processing methods, silica has formed the foundation for a plethora of technologies (21, 73–81). However, in the context of nonlinear performance, one significant limitation of silica is its low nonlinear behavior. Unless in the form of quartz (which is rare in an integrated platform), silica is an amorphous material, lacking inherent symmetry. While methods like high temperature or thermal poling can be used to induce an electro-optic ($\chi^{(2)}$) response, it is adds complexity to fabrication (82). However, due to the exceptionally low optical loss of silica, it is possible to fabricate devices allowing optical field amplification without damaging the material, allowing nonlinear behaviors to be excited (83–93). In addition, the ease of doping the amorphous silica matrix provides a complementary route to designing and realizing nonlinear devices from silica (18, 77, 84, 94–99).

Silicon nitride is a popular wide bandgap dielectric material that has been the focus of numerous research efforts. The measured single mode propagation loss for silicon nitride is below 1 dB/cm in the 532–900 nm range (100). Due to the central-symmetric crystalline quality, its $\chi^{(2)}$ nonlinearity is zero except under the presence of interface strain (58) and strong bias (101). The $\chi^{(3)}$ nonlinearity of silicon nitride in the form of Kerr refractive index ($n_2$) was identified to be $2.5\times10^{-15}$ cm$^2$/W at 1550 nm by evaluating the nonlinear optical response of an optical resonator (102). It was also reported that the nonlinear response can be further enhanced by 5 times if silicon-rich silicon nitride waveguides were adapted (103). One of the major advantages of silicon nitride platform is its excellent CMOS compatibility (104, 105), which provides the potential for multi-layer (106) and high-density integration (107).

In addition to silica and silicon nitride, several other materials have been studied for nonlinear applications in the visible spectrum. For example, lithium niobate has been extensively investigated due to its strong $\chi^{(2)}$ on the order of $d_{33}$ = −20.6 pm/V (108) and the significant electro-optic effect (109). Recent developments on Lithium-niobate-on-insulator (LNOI) technology (110, 111) showed high potential for its adaption in integrated photonic applications. An ultra-low loss of 6 dB/m at 637 nm has been achieved (110) in a lithium niobate waveguide with visible electro-optic bandwidth exceeding 10 GHz. In addition, a recent work showed that 4H-silicon-carbide-on-insulator (4H-SiCOI) technologies (112) offer unique optical properties for quantum photonic and nonlinear optical applications, where a relatively high second order nonlinearity of $d_{33}$ = −12.5 pm/V at 1.064 μm (113) is observed. This is promising for efficient parametric conversions. However, further investigation is still required for this new platform. Another promising material is diamond. It has excellent optical properties for quantum emitters (114) and PICs (115), and it has been proposed for wide variety of nonlinear optical applications. Atomic layer deposited $Al_2O_3$ was also reported with excellent wave guiding behavior in the UV spectrum with propagation loss of < 3dB/cm at 371 nm (116); however, it has a relatively weak modal confinement and a small Kerr refractive index (117). AlGaAs possesses strong third order nonlinearities ($n_2$ = 2.6×10$^{-13}$ cm$^2$/W) over most popular dielectric materials, which leads to Kerr comb generation with extreme low threshold values (118). Comprehensive reviews and representative demonstrations on aforementioned materials can be found in the prior references. Due to the space limit, we focus our discussions on GaN-based III-N and organic materials, which have emerged as two groups of exciting new materials for nonlinear integrated photonics applications particularly for UV–Vis wavelength range.

## 3.2   III-N materials

The overall advantages of III-N materials originate from their wide bandgap energies, outstanding properties in both optoelectronics and nonlinear optics, and excellent compatibility with existing III-N light sources in the UV–Vis spectral range. The properties of GaN includes a wide bandgap energy of 3.4 eV, corresponding to a transparent wavelength above 365 nm. Due to the close lattice constant with InGaN LEDs/lasers (119) and AlGaN RF electronic components (120), GaN is promising for active opto-electronic integration. By Maker-Fringe measurements, the quadratic nonlinear-optical coefficient of GaN was identified to be $d_{31}$ = 2.5±0.1 pm/V and $d_{33}$ = −3.8±0.1 pm/V (121). Some early characterizations on nonlinear optical properties of GaN layers on sapphire substrates (i.e., typically with defect density of >10$^9$ cm$^{-2}$) using the Z-scan method suggests a two-photon absorption (TPA) coefficient of β = 17±7 cm/GW at λ = 400 nm and β = 3±1.5 cm/GW at λ = 720 nm (122). In contrast, recent results on GaN bulk materials (i.e., with threading dislocation defect density of <10$^6$ cm$^{-2}$) reveal that the intrinsic TPA coefficient is about 0.9 cm/GW at 724 nm (123) and 3.5 cm/GW at 532 nm (124). The improvements in the TPA coefficients on GaN bulk materials can be attributed to improved material quality, by which the density of deep level defects (125) and n-type donors (nitrogen vacancies) (126) are greatly reduced compared to GaN layers on sapphire substrates. The Kerr refractive indexes ($n_2$) of GaN were identified to be $n_2$ = 1.15–1.4×10$^{-14}$ cm$^2$W$^{-1}$ at λ = 800 nm (127). However, because the growth of GaN is usually accompanied with large amount of threading dislocations, the two-photon absorption coefficient of GaN is relatively large in comparison with other wide bandgap semiconductors (128).

Compared to GaN, AlN has an even wider bandgap energy of 6 eV, allowing broadband transparency from deep-UV to infrared. The quadratic nonlinear optical coefficient of AlN was characterized by A. Majkić *et al*. (129). This work measured an absolute value of $d_{33}$ = 4.3±0.3 pm/V at λ = 1030 nm, and the $d_{31}$ was in the range of $d_{33}$/(45±5). Since the AlN is mainly grown on *c*-plane of sapphire substrate, the measured $d_{33}$ and $d_{31}$ suggests that TM modes possess a larger second order nonlinearity, which will lead to a higher conversion efficiency in second harmonic generation (SHG) (130) and a degenerate parametric down conversion (131). Due to the difficulties in obtaining high quality bulk AlN and the large two-photon absorptive photon energy, direct Z-scan measurement on high quality AlN is still not reported. One exception is the report by M. Zhao *et al*. (132), in which the TPA coefficient and $n_2$ at λ = 355 nm were measured to be 13±3 cm/GW and $-1.91\pm0.38\times10^{-13}$ cm$^2$W$^{-1}$, respectively. Noting that the AlN sample used in the previous work (132) exhibits strong optical absorption below λ=300 nm, possibly due to the inferior crystalline quality of the AlN. Alternatively, one can also adopt the derived $n_2$ of 2.3±1.5×10$^{-15}$ cm$^2$W$^{-1}$ at 1550 nm for AlN (133), and utilize the wavelength dependence fitting provided in previous work (134) to roughly estimate the nonlinear optical performance of AlN in the UV and visible spectral range. The estimated $n_2$ at 800 nm is 3×10$^{-15}$ cm$^2$W$^{-1}$ and 1×10$^{-14}$ cm$^2$W$^{-1}$ at 400 nm.

Table 2: Nonlinear optical properties of III-N in comparison with other materials.

| Material | Optical bandgap (eV) | Refractive index | Dominant $d_{ii}$ (pm/V) | $n_2$ (cm$^2$/W) | Linear loss in visible (dB/cm) |
|---|---|---|---|---|---|
| SiO$_2$ | 8.9 | 1.45 | \ | 4×10$^{-16}$ (135) | 0.97 (at 658 nm) (136) |
| Si$_3$N$_4$ | 5.1 | 2.05 | \ | 2.5×10$^{-15}$ (102) | < 1 (532 ~ 900 nm) (100) |
| SiC | 2.3 ~ 3.3 | 2.6 | −12.5 (113) | 1.9±0.7×10$^{-15}$ (137) | 2.3 (at 633 nm) (138) |
| Al$_2$O$_3$ | 9.1 | 1.77 | \ | (3.3 ~ 2.8)×10$^{-16}$ (117) | < 2 (at 405 nm) (116) |
| Diamond | 5.5 | 2.42 | \ | 1.3×10$^{-15}$ (115) | \ |
| AlGaAs | 1.4 ~ 2.2 | 2.9 ~ 3.4 | 90 (139) | 2.6×10$^{-13}$ (118) | \ |
| TiO$_2$ | 3 ~ 3.5 | 2.4 | \ | 2.3~3.6×10$^{-14}$ (140) | 28 (at 635 nm) (141) |
| LiNbO$_3$ | 3.7 | 2.2 | −20.6 (108) | 5.3×10$^{-15}$ (142) | 0.06 (at 637 nm) (110) |
| GaN | 3.4 | 2.4 | −3.8±0.1 (121) | 1.15 ~ 1.4×10$^{-14}$ (127) | ~ 2 (at 700 nm) (143) |
| AlN | 6.2 | 2.16 | 4.3±0.3 (129) | 2.3±1.5×10$^{-15}$ (133) | 3±1 (at 635 nm)(144) |

## 3.3 Organic materials

While initial work in the field of organic photonics focused on polymers (145, 146), organic small molecules are emerging as an alternative material for a wide range of applications (147–153). Organic nonlinear optical small molecules are π-conjugated molecules, and the strong nonlinear optical susceptibilities and rapid responses are based on highly movable π-electrons along the molecular backbone(154–158). While the timescales of the nonlinear response are comparable, the magnitudes that can be achieved are larger than in conventional crystalline systems, enabling higher performing devices. In addition, the nanometer-scale length or size of the material needed to achieve a nonlinear response is much smaller than either conventional crystalline materials or organic polymers. However, the response is dependent on the molecular orientation of the molecule.

Specifically, the nonlinear optical response of an organic molecule that is induced can be described by the expression below (159):

$$\mu_i = \alpha_{ij}E_j + \beta_{ijk}E_jE_k + \gamma_{ijkl}E_jE_kE_l + ... \qquad (18)$$

where $\mu_i$ is the induced dipole moment in a molecule, $E_i$ is the electromagnetic field, α, β, and γ are the linear polarizability, first hyperpolarizability, and second hyperpolarizability tensors, respectively. It is important to note that this expression describes the efficiency of charge transfer (β) or vibration (γ) at the molecular level (160). The $\beta_{ijk}$ is responsible for second order nonlinear effects, and the $\gamma_{ijkl}$ is responsible for third order nonlinear effects, as described by the macroscopic susceptibilities $\chi^{(2)}$ and $\chi^{(3)}$ in Eq 6. However, it is important to note that the macroscopic susceptibilities are related, but not equal, to the molecular nonlinearities. Several factors influence this relationship including the density of the organic molecules and the orientation of the molecule with the incident optical field. As one might imagine, there are nearly limitless possible organic chemical structures that could be synthesized that would exhibit nonlinear behavior.

For comparison, Table 3 lists $n_2$ of a variety of inorganic and organic materials mentioned in this review. In inorganic materials conventionally used to fabricate integrated optical circuits, $n_2$ follows the range of ~$10^{-18}$ to ~$10^{-20}$ m²/W. Organic materials show a higher magnitude but similar variance of $n_2$ values. However, one challenge when making a table is determining the absolute value to report. As mentioned, the $n_2$ depends on the molecular density and molecular orientation with the optical field.

Table 3: Nonlinear optical properties of several commonly used organic materials.

| Material | Structure | Optical bandgap (eV) | Refractive index | First-order hyperpolarizability (m/V) | $n_2$ (cm²/W) |
|---|---|---|---|---|---|
| DASP | 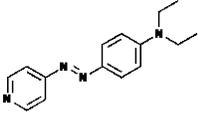 | 3.9** | 1.57 | $(2.6 \sim 3.7) \times 10^{-32}$ (161) | $2.5 \times 10^{-13}$ (162) |
| TPE | 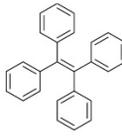 | 3.6 (163)* | 1.64 | $3.1 \times 10^{-33}$ (163)* | $1.9 \times 10^{-15}$ (164)* |
| Anthracene | 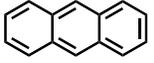 | 3.1 (165) | 1.59 | $2.2 \times 10^{-37}$ (166) | $2.6 \times 10^{-15}$ (167) |
| PTCDA | 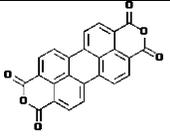 | 2.1 (168) | 2.21 | \ | $1.2 \times 10^{-13}$ (169) |

*Values of derivative structures. ** Calculated by time-dependent density functional theory with the gas phase ground state molecular geometry optimized at the B3LYP/6-31G* level of theory

To limit the scope of this article, we will focus on the two different chemical structures shown in Figure 5: Tetraphenylethylene (TPE) and 4-[4-diethylamino(styryl)]pyridinium (DASP).

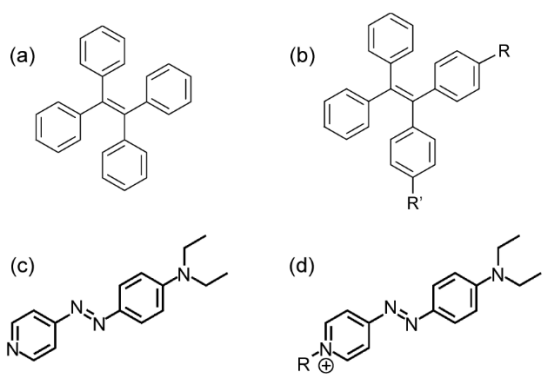

Figure 5: Chemical structures of commonly used organic small molecules. (a) TPE and (b) TPE with common sites for modification indicated by R and R'. (c) DASP and (d) DASP with common sites for modification indicated by R.

TPE is an intriguing material with a unique chemical structure. In previous work, it has demonstrated large NLO coefficients, particularly the first hyperpolarizability or the second-order NLO coefficient, and the four-leaf clover architecture facilitates the design of push-pull chromophores that allows electron charge transfer across the π conjugation of the TPE molecule (170–179). It has previously been used successfully as an imaging agent (155, 155, 180), and, in contrast to many fluorophores, it is well-known for exhibiting aggregation-induced-emission (AIE) (171, 181) in which the emission intensity increases as the molecules aggregate.

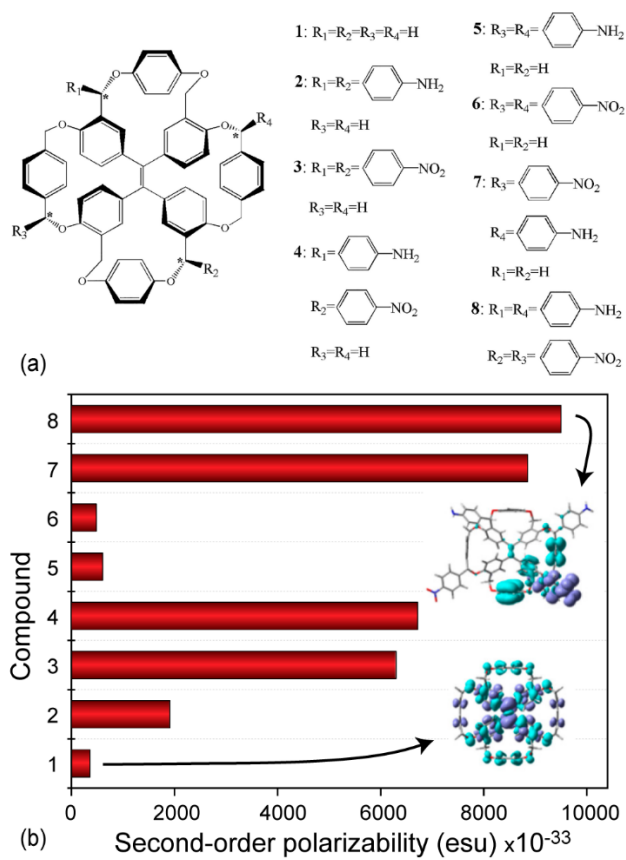

Figure 6: (a) Structures of a series of cyclic TPEs. (b) Second-order polarizabilities of the synthesized TPE derivatives in part (a). The second order polarizability increased approximately 27 times from $1.50 \times 10^{-34}$ m/V to $3.98 \times 10^{-33}$ m/V (converted from esu units) by changing the residues. Inset: DFT modeling of compounds 1 and 8. Adapted with permission from C. Liu, G. Yang, Y. Si, X. Pan, Journal of Physical Chemistry C. 122(9):5032–39 (2018). Copyright 2018 American Chemical Society

One study measured the nonlinear optical activity of a series of TPE structures (Figure 6) (170). In this study of eight different structures, the second-order NLO polarizabilities was increased as much as 27 times by tuning the substituent groups and their positions even when the primary chemical composition (the TPE core) was relatively unchanged. As can be seen in Figure 6, Structure 1 starts with hydrogens in all $R_1$-$R_4$ substituent groups. In this case, the electron density distribution of Structure 1 from the density functional theory (DFT) calculation is fairly uniform (Figure 6b, inset), meaning that the intramolecular charge transfer (ICT) is less likely to occur. In contrast, in Structure 8, the $R_1$ and $R_4$ were replaced by the donor Ph-$NH_2$, and $R_2$ and $R_3$ were replaced by the acceptor Ph-$NO_2$. In this molecule, the electron density distribution is highly asymmetric due to the asymmetric orientation of the donor-acceptor R groups (Figure 6b, inset). It is important to note that this work focused on tuning the conjugation length via $NH_2$ and $NO_2$ groups. However, there are numerous other strategies that have also been employed successfully, such as bromination.

In addition to purely synthetic routes, researchers have also employed predictive methods. These have been particularly successful in accelerating material design for small organic molecules which have fewer atoms than larger polymeric systems. Initial approaches, such as DFT, were based on quantum mechanical calculations of the orbital structures (182–185). Hydrogen bond pattern prediction based on similar quantum mechanical modeling has also been successful in the design and prediction of NLO organic molecules. Freely available shareware has popularized these methods among chemists and material scientists impacting material design in a range of fields (186–194). More recent computational methods have relied on utilizing an algorithm which can quickly find minimums in the potential energy landscapes of novel organic materials.

The chemical structures of 4-[4-diethylamino(styryl)]pyridine (DASPy) and 4-[4-diethylamino(styryl)]pyridinium (DASP) are depicted in Figure 5. When the pyridine moiety is substituted with a R group, DASPy becomes its pyridinium form DASP. As seen, there is one equivalent of positive charge on pyridinium in DASP compared to DASPy, and this additional formal charge makes DASP a "push-pull" chromophore where electron-rich amine group favors to donate electrons to electron-poor pyridinium (195–198). Thus, as long as DASP molecules can be functionalized wisely on a specific surface, such delocalization of electrons on DASP makes it suitable for photoactive materials in NLO studies.

Over two decades ago, scientists began investigating the SHG behavior in DASP Langmuir-Blodgett (LB) films. To further improve the nonlinearity, one reported a series of rare earth metal dysprosium-DASP complexes(199, 200). According to the authors, dysprosium complexes serving as the R group facilitate the formation of uniform LB films. They characterized the SHG with the Nd:YAG laser beam at with a peak power at 1064 nm, anticipating SHG emissions at 532 nm. Both of the compounds investigated achieved $\chi^{(2)}$ and $\beta$ values in the $10^{-6}$esu and $10^{-27}$esu range, respectively. Zhao *et. al*. also discovered that the solvent environment has an effect on SHG performances of DASP LB films(201). For example, when the vapor pressure of isopropanol increases, the SHG signal decreases. Even after the solvent vapor pressure is removed, the SHG signal is not fully recovered. Such considerations will play a key role when considering and optimizing processing conditions for integration with on-chip optical devices.

## 4    Device Platform and Applications

### 4.1    Device platforms

The most straightforward device to fabricate on a silicon wafer is an integrated waveguide. The first on-chip waveguides were simple slab, or rectangular, waveguides with uniform indices (202). These devices essentially mimicked optical fibers on-chip. Subsequently, researchers realized that nanofabrication opened up possibilities not easily available when drawing fibers, including index and geometric variations over nano-scale dimensions (203–205). Such devices enable control over the material dispersion, and therefore, they form a key component in the engineer's toolbox, enabling nonlinear behaviors to be accessed.

A second commonly used device is the optical resonant cavity. Both traveling wave and standing wave resonators have been used in nonlinear optics investigations as both cavities are able to achieve the high optical intensities needed to unlock nonlinear behaviors (206). Photonic crystal cavities, a common standing wave resonator, have small optical mode volumes which focus the light, resulting in high intensity (41). In contrast, while whispering gallery mode cavities have large optical mode volumes, they can achieve high optical cavity quality factors (Q) due to their long photon lifetimes (83). The high Q values allow them to achieve large circulating intensities.

Initial work investigating nonlinear phenomena was in the near-IR and focused on leveraging the ultra-high-Q factors possible in the whispering gallery mode cavities, particularly in silica and fluoride devices (207). Despite the low nonlinear coefficients, Stokes and Anti-Stokes behavior (208–210) as well as four wave mixing (21, 85, 88, 99, 211–213) were demonstrated. More recent efforts which are the focus of this section are investigating new material platforms to improve the nonlinear coefficient of the material and to open the door to new wavelength ranges.

## 4.2 III-N materials

The major nonlinear optical application of GaN is the SHG from the telecommunications wavelengths (~1550 nm) to the near-visible spectral. C. Xiong *et al.* reported the SHG from GaN ring resonators with measured quality factors around 10,000 at ~1550 nm (214). I. Roland *et al.* also reported the SHG at similar wavelengths using GaN-on-silicon suspended disk resonators (215), and the typical loaded *Q* factors were in the range of 6,000 to 13,000. However, the typical *Q* factors of GaN resonators are limited at $10^4$, which are more than two orders of magnitudes lower than the typically *Q* values obtained from resonators based on other materials (216, 217) with similar dry etch fabrication methods. Recently, H. Chen *et al.* investigated the loss mechanisms on GaN photonic devices and suggested that the nitrogen vacancy contributed n-type conductivity ($N_0 = 10^{18}$ cm$^{-3}$) was responsible for the high propagation loss in GaN devices, where a high free carrier absorption loss of > 2 dB/cm be expected (143). This high degree of free carrier absorption has hindered the development of high *Q* resonators for GaN devices, which are typically epitaxially grown and fabricated on sapphire or silicon substrate and therefore contain high defect densities. It is noteworthy that the recent development of GaN devices fabricated on high quality GaN bulk substrates (134) could potentially tackle this challenge by reducing the density of nitrogen vacancies in the GaN materials. Further investigations on these new "GaN-on-GaN" devices are required to achieve high *Q* GaN resonators.

Compared to GaN, the wider bandgap of AlN allows for a very low n-type conductivity, and consequently much lower free carrier absorption loss. As a result, extensive investigations have been carried out on AlN devices, where AlN ring resonators with very high *Q* values have been demonstrated. In the telecommunications spectral region, the AlN ring resonators showed typical intrinsic *Q* factors on the order of $10^6$ for both TE and TM modes (216, 218). In the near-visible spectral region, Y. Sun *et al.* demonstrated AlN ring resonators with intrinsic *Q* factors greater than 70,000 (219), while T. J. Lu *et al.* reported AlN devices with intrinsic *Q* values exceeding 170,000 at 638 nm (220). In the UV spectra region, X. Liu *et al.* demonstrated AlN devices with intrinsic *Q* factors of 210,000 (221). AlGaN/AlN alloys were also theoretically proposed (222), in which the low index contrast is promising for low loss waveguiding. The relative higher *Q* factors on AlN devices allows more efficient frequency conversion from telecommunications to near-visible spectrum (223) and vice versa(131) through the $\chi^{(2)}$ processes.

Despite the successful demonstration of high *Q* resonators on AlN, the large chromatic dispersion and group velocity dispersion in the UV and Vis spectra remain to be the significant challenges towards realizing highly efficient frequency conversion through $\chi^{(2)}$ or $\chi^{(3)}$ process using AlN devices. Recently, X. Liu *et al.* proposed a chirp-modulated taper AlN waveguide to achieve broad phase matching using a near-visible pumping wavelength (221). By fully utilizing the $\chi^{(2)}$ and $\chi^{(3)}$ nonlinearities, over 100 THz of coherent frequency spanning (360-425 nm) was achieved on AlN, and the results are as shown in Figures 7a and 7b. Alternatively, coherent generation in the UV and visible spectra can also be achieved using the soliton dynamics. For example, D. Y. Oh *et al.* successfully demonstrated the coherent generation in the UV wavelengths using dispersive wave generation from silica ridge waveguides (224), which is shown in Figure 7c. The dispersive waves were widely tunable from 322 nm to 545 nm by tuning the waveguide geometry. Excellent coherence was maintained since the frequency conversion is phase matched. More recently, H. Chen *et al.* demonstrated that the same engineering principle can be applied to AlN devices, where supercontinuum generation from near UV–Vis spectra was achieved using dispersion engineered AlN waveguides (as shown in Figures 7d and 7e) (225). The AlN waveguides were pumped in $TE_{10}$ mode near the zero dispersion wavelength (ZDW) at 810 nm(225). Due to the relatively large nonlinear parameter of AlN (221, 225) compared with silica (224), the pulse energy of AlN waveguides was significantly reduced to less than 1 nJ while covering a much broader spectral region, while the typical operation pulse energy within silica is above 1nJ (224). This work opens door for integrated AlN nanophotonic devices and circuits, which will lead to potential applications in on-chip mode locking (226), parametric oscillator (227), and entangled photon generation (228).

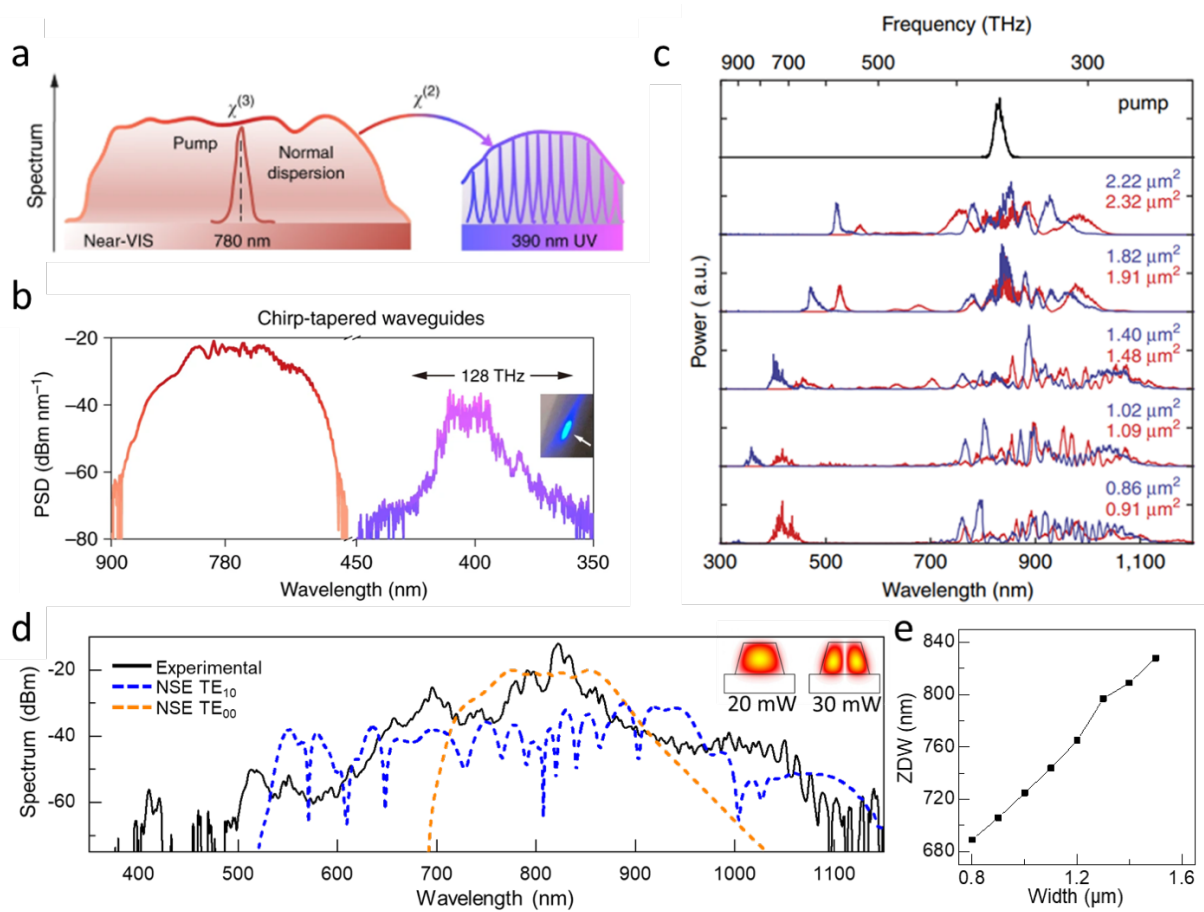

Figure 7. (a) The coherent generation in the UV spectra via $\chi^{(2)}$ and $\chi^{(3)}$ processes simultaneously from a chirp-modulated taper AlN waveguide. (b) The broad spectrum obtained from near-the chirp-tapered AlN waveguide using a near-visible pumping wavelength. Reprinted from (221). Distributed under a Creative Commons Attribution NonCommercial License 4.0 (CC BY-NC) http://creativecommons.org/licenses/by-nc/4.0/ (c) The coherent dispersive wave generation in the UV and visible spectra through soliton dynamics using silica ridge waveguides. Red and blue curves indicate TE and TM mode operation, respectively. Reprinted from (224). Distributed under a Creative Commons Attribution NonCommercial License 4.0 (CC BY-NC) http://creativecommons.org/licenses/by-nc/4.0/ (d) The supercontinuum generation from AlN waveguide pumped near zero dispersion wavelength in $TE_{00}$ and $TE_{10}$ modes. (e) The geometry dependence of ZDW for $TE_{10}$ mode. Anomalous dispersion can be obtained above the ZDW. Reprinted from (225). Distributed under a Creative Commons Attribution NonCommercial License 4.0 (CC BY-NC) http://creativecommons.org/licenses/by-nc/4.0/

### 4.3 Organic materials

The fabrication of devices from nonlinear organic materials is intrinsically more complex than from III-V or II-IV materials due to their inherent incompatibility with the majority of nanofabrication procedures. In past work using organic polymeric materials, it was common to fabricate the entire device from the polymer using either nanoimprinting or laser writing methods (13, 229). Using this approach, several different on-chip structures were fabricated including waveguides, resonators, and modulators with GHz rates(8–16). However, because the device is fabricated from the polymer, the optical performance is limited by the optical loss of the material which can be high. Additionally, because the optical performance is dependent on the orientation of the polymer, the device has to the poled before use, and the effect of the poling on the polymer has a finite lifetime.

To overcome this barrier, researchers began investigating alternative strategies, such as dip-coating or spray-coating polymeric materials on the surfaces of integrated devices (19, 230, 231). Because thin layers could be demonstrated, the negative effect of the polymer's loss was less, allowing higher quality factors to be achieved. Initial demonstrations focused on tailoring the thermal stability of the device, but subsequent investigations demonstrated Raman lasing behavior, albeit with very high thresholds due to the decreased quality factors because of the high loss of the polymer layer (19, 230, 231). While successful, this fabrication approach creates devices where the polymer layer is disordered. However, molecular orientation, both with respect to the optical field and with respect to adjacent molecules, can govern the performance of the resulting device. Therefore, this strategy places fundamental limits on the types of nonlinear behaviors that can be accessed as well as the efficiencies and thresholds possible.

To realize the potential of small organic molecules in integrated photonics, fabrication methods which allowed oriented and ordered monolayers of molecules had to be developed. One such route was inspired by earlier work in the biosensor field (232–238). By self-assembling and covalently attaching nonlinear organic molecules onto the device surface, these criteria can be achieved.

One example demonstration was based on combining a silica cavity with a monolayer of DASP molecules (232). This strategy leveraged the intrinsic hydroxyl groups present on a silica device surface as anchor sites for the highly nonlinear DASP molecules, thus allowing all aspects of the molecular orientation to be controlled (Figure 8a). However, because the circulating optical field is primary confined within the silica device, the net effect of the molecule on the efficiency will be governed by a weighted $\chi^{(3)}$, described by the relative overlap of the optical field in each material. Despite the reduced effect, the DASP molecule is still able to significantly increase the performance (Figures 8b and 8c)). In complementary work that leverages the same surface chemistry and optical cavity type, a DASP-based surface coating is used to demonstrate third harmonic generation with approximately a 4 orders of magnitude improvement in efficiency over a non-functionalized device (90). This strategy allowed a near-IR laser to be used to reach the visible. Both of these examples relied on the nonlinearity present in the molecule to create a nonlinear device.

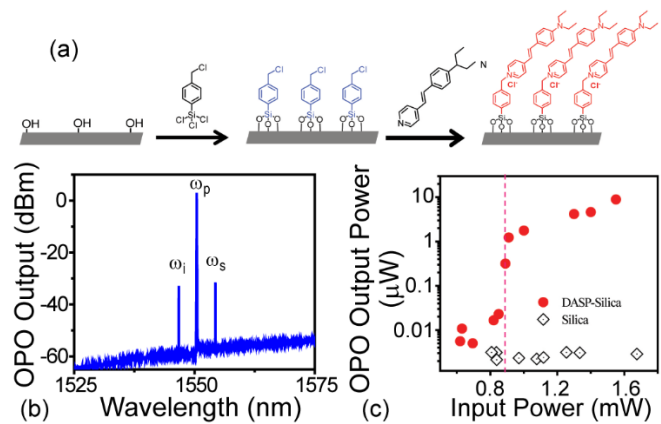

Figure 8: DASP coated silica resonant cavity for four wavemixing. (a) Overview of surface chemistry method used to create an aligned monolayer of DASP molecules on the device surface. (b) Representative optical parametric oscillation (OPO) spectrum. (c) Threshold power comparing non-functionalized and DASP-functionalized devices. Reproduced from reference (232). Distributed under a Creative Commons Attribution NonCommercial License 4.0 (CC BY-NC) http://creativecommons.org/licenses/by-nc/4.0/

Recent work demonstrated that by perturbing the surface of a device, it is possible to improve the device's nonlinear performance. Specifically, by attaching an oriented siloxane monolayer (Si-O-Si) on a silica surface, the ability of the underlying device to generate stimulated Raman scattering or create a Raman laser is enhanced due to the creation of a surface Raman mode in this oriented and ordered layer. As expected, due to the ordering, the threshold and efficiencies of this mode are polarization dependent, unlike in the amorphous device. Raman (Stokes) is a particularly flexible excitation mechanism because it is not dependent on a specific electronic transition. As such, this approach for increasing the efficiency of Stokes and Anti-Stokes generation could provide an accelerated path towards UV emission.

In the near future, it is expected that the harmonic signals can reach the UV range by precisely tuning the organic NLO materials and the frequencies/wavelengths among the pump, Raman Stokes, and parametric oscillation photons. The innovations of photonic devices are also probable as nanofabrication allows more options to design on-chip photonic devices as long as the functionalization of organic NLO materials are feasible.

## 5   Future Directions

### 5.1   New materials

Recently, a new class of wide band gap semiconductor beta-phase gallium oxide (β-$Ga_2O_3$) has emerged with many promising properties. β-$Ga_2O_3$ possesses a bandgap energy of 4.8 eV and exhibits large laser induced damage threshold (LIDT) (239), which is promising for high power PICs. The low refractive index contrast between core and cladding layers also minimize the scattering loss (240). Furthermore, β-$Ga_2O_3$ also has a small lattice mismatch with the InAlGaN material system, which leads to possible active integration of $Ga_2O_3$ photonic devices with InAlGaN lasers and detectors.

Preliminary efforts have been paid towards the fundamental optical properties as well as device performance of β-$Ga_2O_3$. H. Chen *et al.* characterized the optical nonlinearity of β-$Ga_2O_3$, where a TPA coefficient of 0.6–3.2 cm/GW, and an $n_2$ in the range of $-2.1\times10^{-15}$ $cm^2W^{-1}$ to $-2.9\times10^{-15}$ $cm^2W^{-1}$ were obtained at 404 nm (128). It's noteworthy that bulk crystalline β-$Ga_2O_3$ was obtained using the floating zone growth method (241), which leads to much superior crystal quality with less defect density on β-$Ga_2O_3$ compared to GaN bulk crystals. As a result, the TPA coefficient of β-$Ga_2O_3$ is lower than GaN (see Figures 9a and 9b). More recently, J. Zhou *et al.* successfully fabricated the β-$Ga_2O_3$ waveguides and analyzed their performance in the UV to near infrared (NIR) region (240) (see Figures 9c and 9d) (240). A low propagation loss of 3.7 dB/cm was obtained on the β-$Ga_2O_3$ waveguide at the wavelength of 810 nm, which is comparable to the state of the art. Combined with theoretical simulations, various loss mechanisms from two-photon absorption, sidewall scattering, top surface scattering, and bulk scattering were discussed for β-$Ga_2O_3$ waveguides, and their contributions to the total optical loss were estimated. It is expected that the performance of β-$Ga_2O_3$ photonic devices could be greatly improved with the further development in materials synthesis, optical design, and device fabrications, which will unleash the full potential of β-Ga₂O₃ for UV–Vis photonics applications.

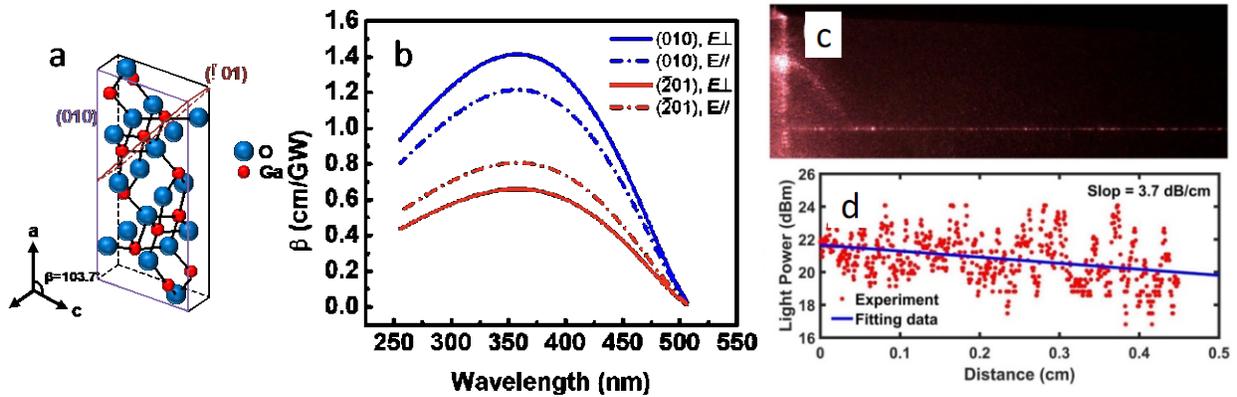

Figure 9: (a) The schematic of β-Ga₂O₃ crystal structures showing (010) and ($\bar{2}$01) crystal planes (128). (b) The estimated wavelength dependence of TPA coefficient for (010) and ($\bar{2}$01) β-Ga₂O₃ samples. "E⊥" indicates that the electrical field intensity is perpendicular to [102] direction, while "E ∥" indicates that field intensity is parallel to [102] direction. Reprinted with permission from (128). Copyright 2018 by the Optical Society of America. (c) Top image captured by the linear CMOS camera of a β-Ga₂O₃ waveguide with 1.5 µm width at 810 nm wavelength. (d) Experimental data and regression analysis of β-Ga₂O₃ waveguide with 3.7 dB/cm loss. Reprinted with permission from (240). Copyright 2019 by the American Physical Society.

In addition to emerging crystalline materials, there is a whole host of new organic materials on the horizon. One simple organic platform is Anthracene (150, 153, 242, 243). This molecule is characterized by its unique structure which consists of three fused benzene rings, giving it a highly conjugated, resonant structure and allowing it to readily pack into a semi-crystalline structure in the condensed phase. It has previously been used as an organic dye and as a scintillator in organic semiconductors. Moreover, because of their extended conjugation length, anthracenes can make for highly nonlinear molecules due to the delocalization of $\pi$ electrons over the extent of the entire molecule (244). This delocalization allows the molecule to become more polarized given an external oscillating field. This higher polarizability gives rise to 2$^{nd}$ and 3$^{rd}$ order phenomenon even at modest field strengths.

Qualitative reports on the nonlinear activity of anthracenes varies widely due to the multitude of unique derivatives which can be synthesized. Basic DFT models on anthracene crystals without additional functional groups report only modest values for $\chi^{(3)}$ (242). In contrast, other theoretical studies report marked enhancement of the first hyperpolarizability, and therefore nonlinear optical properties, when anthracenes are coordinated with halide acids (166). However, studies on oriented, ordered monolayers of anthracene thin films or integration of anthracenes with integrated photonic devices have yet to be performed.

If one considers moving to a slightly more complex molecular structure, experimental reports have shown large enhancements in both second-harmonic generation and third-harmonic generation when anthracene is synthesized into a highly-conjugated one-dimensional crystalline polymer, or metal organic framework (MOF) (245). The developed MOF's ligands possess an acceptor–π–donor–π–acceptor structure which is symmetric and a singlet biradical electronic ground state, thus boosting its $\chi^{(3)}$ and $\chi^{(5)}$ optical nonlinearities. The reported complex's $\chi^{(3)}$ can reach as high as $8\times10^{-11}$ esu, which is roughly three orders of magnitude higher than typical bulk dielectrics such as $SiO_2$ and ZnO (246). When anthracene is attached along the backbone of a methyl methacrylate repeat unit in PMMA thin films, the enhancement in $\chi^{(3)}$ is even larger, reaching values of about $\sim10^{-7}$ esu (243). This is due to the expanded conjugation length, as the polarizable cross-section now spans the length of a polymer chain (~100kDa), allowing for an increased nonlinear response. Reaching $\chi^{(5)}$ optical nonlinearities in classic crystalline materials is typically not experimentally achievable due to the optical intensities that would be required. Thus, organic small molecules truly provide a route to achieve currently unavailable device performance.

Lastly, though not the focus of this work, an emerging area is creating low-power integrated optical systems by leveraging the intrinsic material properties to achieve functionality, such as phase change materials for *Q* switching. Organic materials offer a particularly unique strategy to achieve this goal. Specifically, unlike in conventional devices where layers consist of a single material type, with organics, it is possible to attach more than one type of molecule to a device surface, creating a multi-functional multi-material monolayer. This concept was demonstrated recently using a photo-responsive group (247, 248), but it has yet to be extended into the realm of nonlinear optical devices.

## 5.2 Inverse design to advance device performance

Thus far, this review has focused on the relatively linear advances in new materials and in more complex device architectures. Recent advances in theoretical device design has opened the door for transformational changes.

Nanophotonic devices often span many wavelengths while having feature sizes on the order of a wavelength, implying that neither lumped nor ray optic approximations suffice and they must be modeled via full-wave solution of Maxwell's equation. The difficulty in modeling nanophotonic devices and lack of analytical solutions makes designing new devices especially challenging. Inverse design, which is the automated design of new devices given target specifications and design constraints as input via optimization, has recently emerged and seen considerable success in the nanophotonics field (249). Although many optimization algorithms have been applied successfully towards the inverse design of nanophotonic devices, including Direct Binary Search (250) and Binary Particle Swarm Optimization (251), local gradient-based search methods have become among the most popular and successful approaches largely due to the adjoint method.

The adjoint method, which first appeared in the context of fluid dynamics (252), enables the computation of the full gradient of the objective function being optimized with respect to the optimization variables with just *two* simulations *regardless* of the number of parameters—the forward problem and the adjoint system (253). This is in contrast to naïve approaches such as finite difference approximations which require N+1 simulations per gradient evaluation where N is the number of optimization parameters. The adjoint method has been applied to design many different devices as shown in Figure 10 including wavelength demultiplexers (254, 255), polarization splitters (250), power splitters (251), photonic switches (256), nanophotonic resonators (257), and grating couplers (258). Furthermore, thanks to the flexibility of the gradient-based optimization approach, arbitrary nonlinear objective functions with highly complex constraints can be handled, which has enabled incorporating fabrication (259) and robustness constraints at design time (256, 260).

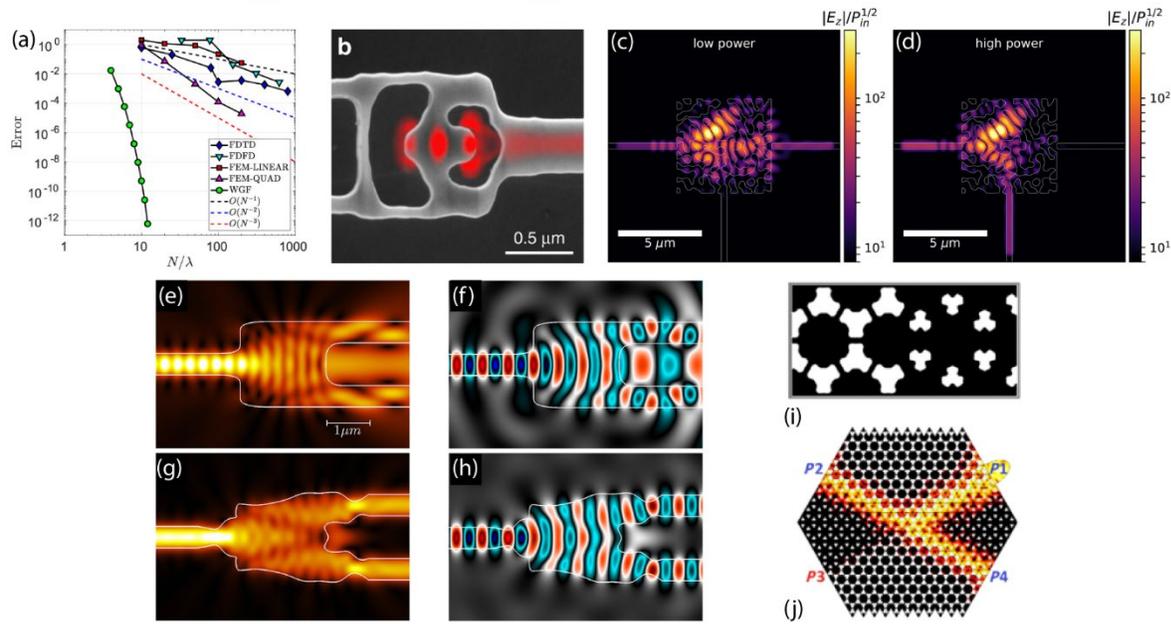

Figure 10: (a) Convergence comparison of BIE, FDTD, FDFD, and FEM solvers with respect to mesh resolution (261). Reprinted with permission from C. Sideris, E. Garza, O. P. Bruno, ACS Photonics. 6(12):3233–40 (2019). Copyright (2019) American Chemical Society. Example devices that have been designed using inverse design include: (b) a vertical coupler on diamond substrate (Reproduced from reference (262). Distributed under a Creative Commons Attribution NonCommercial License 4.0 (CC BY-NC) http://creativecommons.org/licenses/by-nc/4.0/), (c)/(d) nonlinear intensity-based beam splitter (263) (Reprinted with permission from T. W. Hughes, M. Minkov, I.A.D. Williamson, S. Fan, ACS Photonics. 5(12):4781-87 (2018). Copyright (2018) American Chemical Society), (e)-(h) 2x2 power splitter design using BIE solver and adjoint method before and after optimization (261) (Reprinted with permission from C. Sideris, E. Garza, O. P. Bruno, ACS Photonics. 6(12):3233–40 (2019). Copyright (2019) American Chemical Society), (i)/(j) and acoustic topological insulator (264) (Reprinted with permission from R. E. Christiansen, F. Wang, O. Sigmund, Physical Review Letters, 122 (23) (2019). Copyright (2019) by the American Physical Society).

Although traditionally most inverse design implementations in nanophotonics have leveraged the Finite Difference Time Domain (FDTD) method (265) to solve the forward problem, recent work has shown that Boundary Integral Equation (BIE) based techniques can be used to simulate nanophotonic devices up to several orders of magnitude faster and more accurately than finite difference and finite element approaches (261). Furthermore, due to the discretization of global integral operators rather than local difference operators, integral equation methods are essentially dispersion-free, unlike FDTD and finite element method (FEM) which suffer significantly from numerical dispersion. Finally, since BIE methods only discretize the boundaries between different dielectric regions rather than the whole volumes, device optimization via boundary perturbation methods can be implemented naturally and seamlessly without having to resort to level-set approaches (261).

Initial research in inverse design for nanophotonics focused on passive, linear devices on Silicon-on-Insulator (SOI) platforms. Recently, the approach has been extended to time-varying active devices (266–268), topological insulators (264), and nonlinear optics (263), as well as alternative substrates such as diamond (262). In conclusion, inverse design for nanophotonics has demonstrated a strong track record for producing high-performance, compact, and robust devices in numerous different contexts and it has proven to be an invaluable tool for the photonics designer.

# 6 Summary

In summary, the ever-growing materials toolbox provided by emerging semiconductors and small organic molecules optimized for UV-Vis operation will continue to enable device innovation. When coupled with emerging theoretical predictive constructs, such as inverse design for accelerated device design, it is easy to anticipate rapid growth in the field. These new devices will support applications in imaging, cryptography, and spectroscopy as well as enable new fundamental science.

**Acknowledgements**

The authors would like to acknowledge financial support from the Air Force Office of Scientific Research (FA9550-20-1-0087), the Army Research Office (W911NF1810033, W911NF-19-1-0089, W911NF-19-1-0129), and the National Science Foundation (1849965).